# Damage nucleation from repeated dislocation absorption at a grain boundary

Zhiliang Pan, Timothy J. Rupert[*]

Department of Mechanical and Aerospace Engineering, University of California, Irvine, California 92697, USA

*Corresponding Author
Email: trupert@uci.edu
Phone: (949) 824-4937

**Abstract**

Damage nucleation from repeated dislocation absorption at a grain boundary is simulated with molecular dynamics. At the grain boundary-dislocation intersection site, atomic shuffling events determine how the free volume brought by the incoming dislocation is accommodated. This process in turn determines the crack nucleation mechanism, as well as the critical strain and number of dislocations that can be absorbed before cracking. Slower strain rates promote earlier crack nucleation and a damage nucleation mode where cracking is preceded by dislocation emission. The simulation methodology presented here can be used to probe other types of boundaries as well, although multiple thermodynamically equivalent starting configurations are required to quantify the damage resistance of a given grain boundary.

# 1. Introduction

Dislocations can be emitted from, transmitted through, or absorbed by grain boundaries (GBs) in crystalline materials, all of which can lead to damage at the interface. These GB-dislocation interactions are especially important in metals that experience irradiation assisted stress corrosion cracking, where huge numbers of dislocations tunnel through grain interiors and then impinge on GBs [1, 2], in nanocrystalline metals, where dislocation emission and absorption at GBs occurs regularly [3, 4], and in metals under fatigue loading, where dislocations can pile up at a GB [5, 6]. Improving the ductility of structural materials and developing new toughening strategies requires an understanding of damage nucleation mechanisms from GB-dislocation interactions. While existing literature shows that dislocation emission from [7-13] or transmission through [14, 15] different GBs can lead to structural changes [8-11], boundary migration [14], or even fracture of the GB [13], dislocation absorption related damage processes are not as well-understood.

To date, the research that has explored the dislocation absorption process has either simulated a single absorption event [16] or introduced many dislocations of different character by simulating nanoindentation [17] or crack opening [14]. Very little research has focused on damage production from repeated dislocation impingements on a GB. Bitzek et al. [18] showed that the absorption of a single dislocation at nanocrystalline GBs can lead to local stress concentrations, which we hypothesize can then develop into crack damage after repeated absorption. Dewald and Curtin [19] used multi-scale modeling techniques to investigate dislocation pile up at a tilt GB in aluminum, finding evidence of void nucleation. However, the interactions observed by these authors were moderated by significant slip transmission and generation of GB dislocations. In this work, we investigate damage nucleation mechanisms from the complete absorption of multiple dislocations at a GB in face centered cubic (FCC) Cu, using molecular dynamics (MD) simulations. A stable dislocation source was created by mimicking the homogeneous dislocation nucleation process [20] in the center of a bicrystal sample to gradually generate multiple edge dislocation dipoles during loading. We find that dislocation absorption can induce crack nucleation at the GB-dislocation intersection, but this process can be delayed by efficient accommodation of incoming free volume.

## 2. Methods

The bicrystal configuration used in this study is shown in Figure 1(a). The orientation of the center grain (G1) was chosen so that edge dislocation dipoles will propagate to the right and left. The second grain (G2) is oriented such that the resolved shear stress on its slip planes is minimized; therefore, dislocations are entirely absorbed at the GB and the incompatibility between the two grains is not relaxed by direct dislocation transmission into G2. The simulation cell is approximately 61 nm long (X-direction), 32 nm tall (Y-direction), and 9 nm thick (Z-direction), containing ~1,400,000 atoms. MD simulations were performed using the Large-scale Atomic/Molecular Massively Parallel Simulator (LAMMPS) code [21]. Periodic boundary conditions were applied during the simulation and an embedded-atom method (EAM) potential [22] was used to describe the atomic interactions of Cu. All MD simulations were performed at 10 K with an integration time step of 1 fs. The sample was first equilibrated with a conjugate gradient method, and then a Nose-Hoover thermo/barostat was used to further relax the sample for 40 ps under zero pressure.

A positive hydrostatic stress state is usually required to promote crack nucleation and growth in FCC metals [13, 23-25]. This was accomplished here by applying an elastic uniaxial tensile strain of 4% to the sample in the X-direction at a strain rate of $10^9$ s$^{-1}$, in a canonical ensemble. Since Poisson contraction was not allowed during this tensile straining, a positive hydrostatic stress state resulted. After this pre-tension step, the sample was equilibrated for 200 ps using the canonical ensemble. The system reaches equilibrium in less than 20 ps, but we continued the equilibration to access atomistic configurations every 20 ps that are thermodynamically equivalent (relaxed to the same energy, GB structure is identical, etc.) but which differed slightly due to thermal vibrations. This procedure gave us 10 starting configurations, in order to quantify the effect of very subtle thermal vibrations and to allow for increased statistics.

Finally, shear deformation under the canonical ensemble was simulated at engineering shear strain rates of $10^8$ s$^{-1}$ and $10^9$ s$^{-1}$. Due to the technical difficulties of creating an authentic stable Frank-Read dislocation source in atomistic simulations [26], we opted for an artificial method to generate dislocations during our simulations. As shown in Figure 1, we chose two layers of atoms with width of ~5 nm in the center of the sample and displace the layers with respect to each other at a constant speed during the shear step. This displacement is in addition to any movement associated with the overall homogeneous shearing of the simulation cell. The relative speed was chosen so that 10 dislocation dipoles would be created by the time the global engineering shear strain reaches 20%. Although such a source does not occur in real materials, the artificial stress state that results is confined to a region within ~3 nm of the source edge, as shown in Figure 1(b), far enough away from the GBs to not affect any observed GB-dislocation interactions. Two layers of atoms at the bottom of the sample are held fixed in the vertical direction to prevent rigid body grain rotation, which, if allowed, would complicate the simulations. Common neighbor analysis (CNA) was used to identify the local crystal structure of each atom [27], with FCC atoms colored green, hexagonal close packed (HCP) atoms red, body-centered cubic (BCC) atoms blue, and other atoms (usually GB, dislocation, or crack surface atoms) white. All structural analysis and visualization of atomic configurations was performed using the open-source visualization tool OVITO [28].

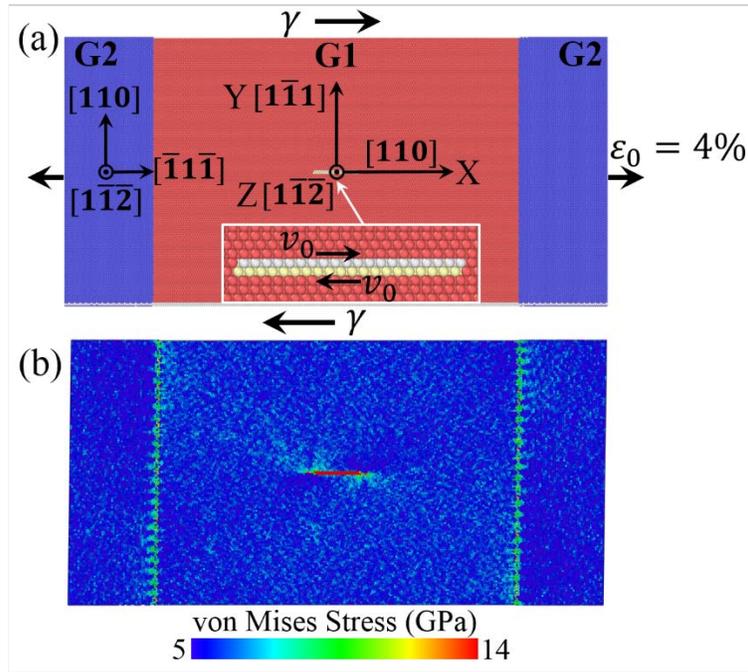

**Figure 1.** (a) Bicrystal atomic configuration containing grains G1 (red atoms) and G2 (blue atoms). Inset shows a dislocation source created in the center of G1 by moving the yellow and white atoms relative to each other at a constant speed. (b) Von Mises stress field at a shear strain of 1.2%, when two leading partials are leaving the source.

## 3. Results and discussions

Intergranular fracture is observed in all samples after multiple dislocation absorption, and two distinct crack nucleation mechanisms can be identified. The first mechanism is shown in Figure 2(a). At a shear strain of 6.4%, after the leading partial of the second dislocation is absorbed, crack nucleation is observed at the intersection of the dislocation with the GB on the left, as shown in the inset. The crack grows and propagates with increasing shear strain, accompanied by the emission of partial dislocations at the crack-tip, as shown in the bottom frame for a shear strain of 6.7%. The second mechanism is shown in Figure 2(b). Here, crack nucleation is preceded by dislocation emission at the GB-dislocation intersection. While two distinct crack nucleation mechanisms are found, crack growth is always accompanied by partial dislocation emission, consistent with observations in the literature [13, 29, 30]. In addition, GB sliding, a very common deformation mechanism for nanocrystalline metals [31], is also observed in all simulations.

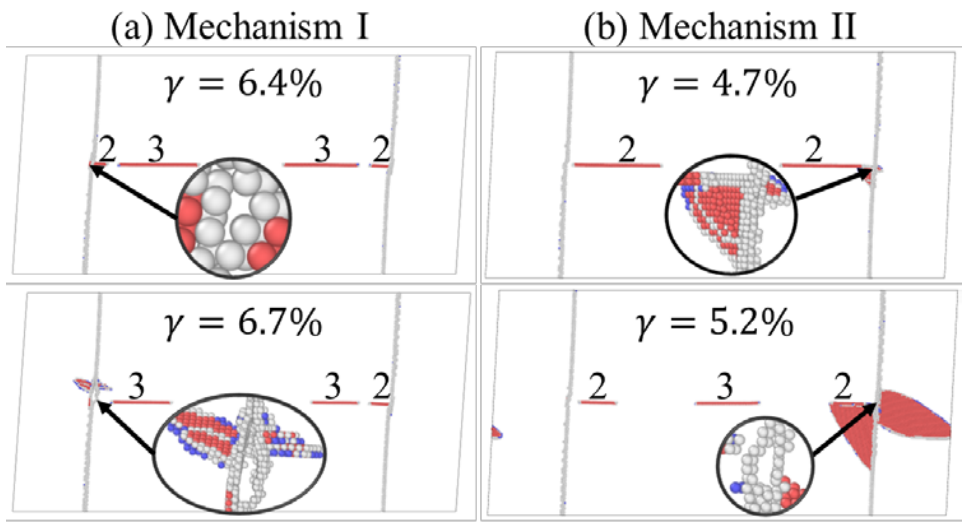

**Figure 2. Two crack nucleation mechanisms observed at the GB-dislocation intersection. Atoms are colored according to CNA, with FCC atoms removed to aid visualization. Insets show the close-up of the intersection. The natural numbers label the sequence of dislocations generated from the artificial source. (a) Direct crack nucleation followed by crack growth. (b) Crack nucleation preceded by dislocation emission.**

The two mechanisms can be connected to atomic shuffling events [12, 17, 31] at the GB-dislocation intersection. Figure 3 shows the close up of the left GB-dislocation intersection of the sample in Figure 2(a). At a shear strain of 2.5%, free volume is formed at the GB after absorbing the first leading partial dislocation. Then, at a shear strain of 3.3%, an atom in the GB jumps into this free volume, preserving the compatibility at the interface. Similar atomic shuffling processes occur through the sample thickness in the Z-direction in all samples that nucleate a crack through Mechanism I. However, if this process does not occur everywhere through the sample thickness and the free volume is only partly occupied, a dislocation is nucleated before crack formation, i.e., Mechanism II commences.

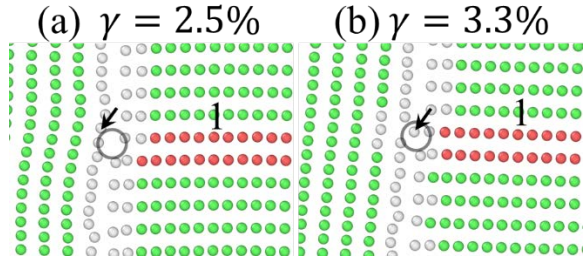

**Figure 3. Close up of the left GB-dislocation intersection of the sample in Figure 2(a) before crack nucleation. Only six atomic layers in Z-direction are chosen and colored according to CNA. (a) Free volume, indicated by the circle, brought by the first leading partial dislocation. An atom from G2 denoted by the arrow jumps to occupy the free volume in G1, as shown in (b).**

To understand how each crack evolves with shear strain and to quantify critical strain for crack nucleation, we plot the crack size as a function of applied shear strain in Figure 5. Here, the crack size is described using the number of atoms sitting on the crack surface, which can be identified using potential energy. As shown in Figure 4, a value of -2.97 eV can successfully identify atoms sitting on a crack surface without finding false positives elsewhere at the interface. Figure 5 shows that the crack starts to grow quickly once the applied shear strain reaches a critical value, which can be used to define the point of crack nucleation. In this work, we define the critical strain for crack nucleation when the number of crack atoms in either GB first exceeds 60, approximately the number of surface atoms for a spherical crack with a diameter of ~1.6 nm. Insets to Figure 5(a) and (b) show a zoomed view of this data and it is clear that the crack size increases quickly at the early stages, indicating that our measurements of critical strain are not overly sensitive to the exact number of crack atoms used to define the nucleation event. At an applied shear strain rate of $10^9$ s$^{-1}$, the average critical strain is 6.3±0.5% for samples which nucleate cracks directly through Mechanism I, higher than that of 5.7±0.3% for samples following Mechanism II. Since Mechanism I represents the process where free volume is fully accommodated by GB restructuring, the correlation between critical strain and mechanism suggests that efficient accommodation of free volume increases the damage resistance of a GB.

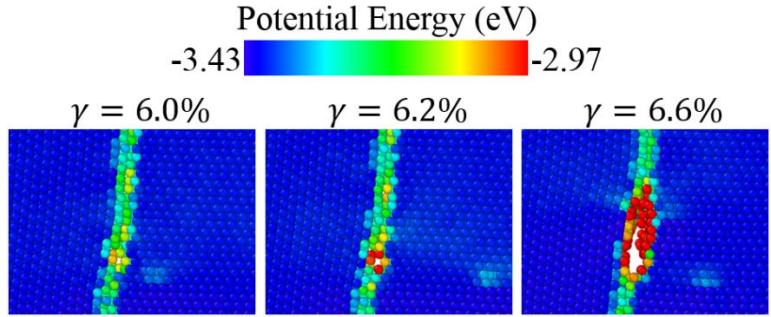

**Figure 4. Potential energy map around a GB-dislocation intersection before and after crack formation.**

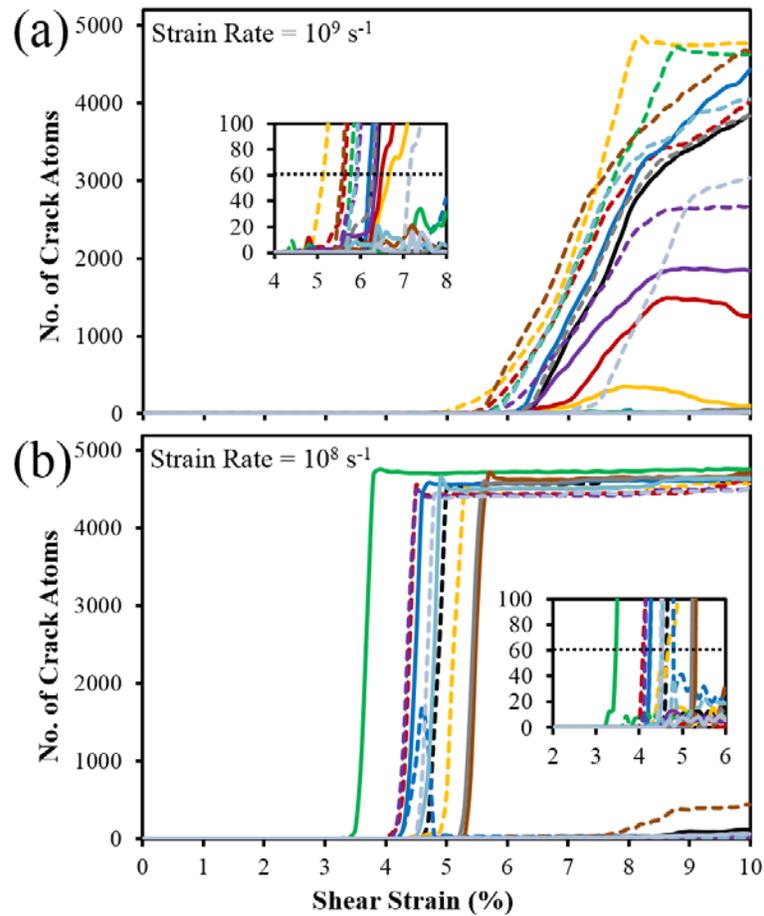

**Figure 5. Number of crack atoms at both GBs as a function of shear strain for strain rates of (a) $10^9\,s^{-1}$ and (b) $10^8\,s^{-1}$. Curves are colored according to starting configurations. Solid curve is for the left GB, while the dashed is for the right. Insets show the crack evolution curve during the nucleation stage, where dotted black lines mark the position of 60 crack atoms.**

The critical strain for crack nucleation and the number of dislocations that are absorbed before cracking, a more physical measurement of a boundary's resistance to crack nucleation that removes any effects of simulation cell size, are summarized in Table 1. The atomic shuffling events which determine the crack nucleation mechanism are thermally activated, giving rise to a strain rate effect; both crack nucleation mechanisms occur at a strain rate of $10^9$ s$^{-1}$, while Mechanism II always occurs at a strain rate of $10^8$ s$^{-1}$. This rate effect can be explained by the fact that a faster strain rate allows less time for the necessary accommodation of GB sliding during the shear deformation, leading to higher local stresses in some cases (depending on thermal vibrations and very subtle details of the boundary structure). The higher local stress that results can in turn reduce the thermal activation energy of the atomic shuffling process, leading to a more military shuffling sequence through the sample thickness and direct crack nucleation (Mechanism I). At the slower strain rate, the local stresses at the interface are lower and shuffling all at once through the simulation cell thickness is unlikely. Instead, accommodation of the free volume occurs in an incomplete manner and Mechanism II commences. While crack size as a function of strain appears to be very different for the two strain rates shown in Figure 5, when the number of crack atoms are plotted versus time, the crack growth rates are very similar. This suggests that crack growth does not have a strong strain rate dependence in our simulations.

The observation of two crack nucleation mechanism demonstrates that thermal fluctuations, even the limited amount that exist at 10 K, can cause noticeable changes in crack nucleation mechanism. However, the variation in critical strain and the number of dislocations absorbed before crack nucleation is marginal, and the methodology used here can measure the resistance of a boundary to damage nucleation from dislocation absorption. For example, although we observe two crack nucleation mechanisms at a strain rate of $10^9$ s$^{-1}$, one and a half dislocations (one full dislocation plus a second leading partial dislocation) are always absorbed before damage forms. The simple perpetual dislocation source presented here makes it possible to evaluate and identify the damage resistance of an individual boundary, and can be applied to any other grain boundaries of interest.

**Table 1. Critical strain and number of dislocations absorbed before crack nucleation, as well as the observed crack nucleation mechanism.**

| Strain Rate | Critical strain | Dislocations absorbed | Nucleation Mechanism |
|---|---|---|---|
| $10^8 \, s^{-1}$ | 4.5±0.6% | 1.15±0.24 | 100% Mech. II |
| $10^9 \, s^{-1}$ | 6.0±0.5% | 1.50±0.00 | 60% Mech. I, 40% Mech. II |

Finally, to elucidate the effect of uniaxial tensile strain (pre-strain), we repeated the simulation of shear deformation of the bicrystal configuration with pre-strain ranging from 1% to 5% at a shear strain rate of $10^9 s^{-1}$. Again, ten nominally identical starting configurations are used in each case. The results presented in Figure 6 show that the number of absorbed dislocations before crack nucleation decreases rapidly with increasing pre-strain. At a pre-strain of 0%, our trend line approaches very large values, suggesting that it would be extremely difficult to initiate a crack in this case. Hence, a nonzero pre-strain is indeed critical for the observation of crack nucleation. The effect of pre-strain on the crack nucleation mechanism is shown in Table 2. Mechanism I (direct crack nucleation) dominates at large pre-strain values, while Mechanism II (dislocation emission and then crack nucleation) dominates at smaller pre-strains. This is consistent with the idea that higher positive hydrostatic stress states favor direct crack nucleation.

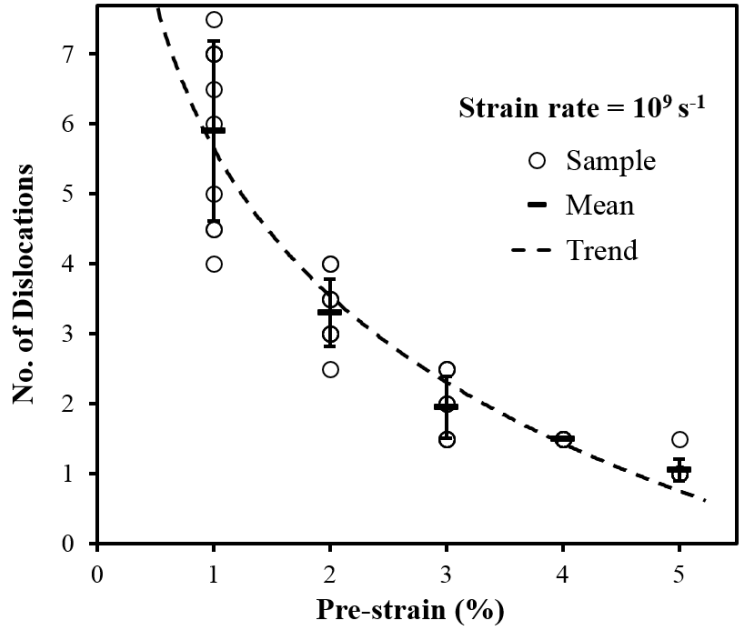

**Figure 6.** Number of absorbed dislocations before crack nucleation as a function of pre-strain.

**Table 2.** The observed crack nucleation mechanism at different uniaxial tensile strains.

| Pre-strain | Nucleation Mechanism |
|---|---|
| 1% | 100% Mech. II |
| 2% | 40% Mech. I, 60% Mech. II |
| 3% | 30% Mech. I, 70% Mech. II |
| 4% | 60% Mech. I, 40% Mech. II |
| 5% | 100% Mech. I |

## 4. Conclusions

An understanding of damage nucleation mechanisms is essential for the development of toughening strategies which can improve the ductility of engineering materials. The MD simulations presented here show that the free volume brought by incoming dislocations must be accommodated at the GB by atomic shuffling events or else a crack forms, suggesting that damage resistant GBs are those that can efficiently absorb free volume due to their local structure. Exactly how the free volume is accommodated determines the crack nucleation mechanism, which is sensitive to thermal fluctuations. However, reliable measurements of a GB's resistance to dislocation absorption-induced damage can still be made based on a statistical average over multiple simulations. Since we observe that free volume accommodation is the key to damage resistance, we propose that GBs with the ability to easily dissolve free volume have the potential to improve the ductility of nanocrystalline metals, reduce the chance of failure from irradiation assisted stress corrosion cracking, and extend fatigue life.

## Acknowledgements

This research was supported by the US Army Research Office through Grant W911NF-12-1-0511.